\documentclass{emulateapj}

\submitted{to appear in {\em Astrophys.\ J.}}

\def\iso#1#2{\mbox{${}^{#2}{\rm #1}$}}
\def\he#1{\iso{He}{#1}}
\def\li#1{\iso{Li}{#1}}
\def\c1#1{\iso{C}{1#1}}
\def\n1#1{\iso{N}{1#1}}
\def\o1#1{\iso{O}{1#1}}

\def\beq{\begin{equation}}
\def\eeq{\end{equation}}
\def\beqar{\begin{eqnarray}}
\def\eeqar{\end{eqnarray}}

\def\la{\mathrel{\mathpalette\fun <}}
\def\ga{\mathrel{\mathpalette\fun >}}
\def\fun#1#2{\lower3.6pt\vbox{\baselineskip0pt\lineskip.9pt
  \ialign{$\mathsurround=0pt#1\hfil##\hfil$\crcr#2\crcr\sim\crcr}}}

\def\ndot{\Gamma}
\def\msol{\mbox{$M_\odot$}}
\def\ndme{{\cal N}_*}

\begin{document}

\title{On Non-Primordial Deuterium Production by Accelerated Particles}

\author{Tijana Prodanovi\'{c} and Brian D. Fields}

\affil{Center for Theoretical Astrophysics,
Department of Astronomy, University of Illinois \\
1002 W. Green St., Urbana IL 61801}

\begin{abstract}
Deuterium plays a crucial role in cosmology because 
the primordial D/H abundance, in the context of 
big bang nucleosynthesis (BBN) theory, yields a precise measure of
the cosmic baryon content.  
Observations of D/H can limit or measure the true
primordial abundance because
D is 
thought to be destroyed by stars and thus
D/H monotonically decreases after BBN.
Recently, however, Mullan \& Linsky have pointed out 
that D arises as a secondary product of neutrons in stellar flares
which then capture on protons via $n+p \rightarrow d + \gamma$,
and that this could dominate over direct D production in flares.
Mullan \& Linsky note that if this process is sufficiently
vigorous in flaring dwarf stars, it could lead to significant
non-BBN D production.
We have considered the production of D in stellar
flares, both directly and by $n$ capture.
We find that for reasonable flare spectra,
$n/d \la 10 $ and $(n+d)/\li6 \la 400$, both of
which indicate that the $n$-capture channel does
not allow for Galactic D production at a level which
will reverse the monotonic decline of D.
We also calculate the 2.22 MeV $\gamma$-ray line production
associated with $n$ capture, and find that
existing COMPTEL limits also rule out significant D
production in the Galaxy today.  
Thus, we find flares in particular, and neutron captures in general,
are not an important Galactic source of D.
On the other hand, we cannot exclude that
flare production might contribute to 
recent FUSE observations of large variations in the local
interstellar D/H abundance; we do, however, offer important constraints
on this possibility.
Finally, since flare stars should inevitably produce
{\em some} $n$-capture events, a search
for diffuse 2.22 MeV $\gamma$-rays 
by INTEGRAL can further constrain (or measure!) 
Galactic deuterium production via $n$-capture.
\end{abstract}


\section{Introduction}

In the past two decades big-bang nucleosynthesis (BBN)
has become one of the most 
important cosmological probes \citep[and refs.\ therein]{osw}. 
Its success lies in the good agreement 
of  predictions for the abundances of four light elements D,$^3$He,$^4$He and 
$^7$Li with their observations. A key element is deuterium,
which is considered to 
be the best cosmic ``baryometer'' among the
light elements \citep{st} because of its strong dependence on 
the baryon density, or equivalently the baryon-to-photon ratio $\eta$.
Furthermore, now that the cosmic microwave background anisotropy measurements
of WMAP have independently measured $\eta$ to 
high precision \citep{wmap}, D takes a new role as a probe of both 
early universe physics and of astrophysics \citep{cfo2,cfo3}.
Therefore, it 
is crucial that we fully understand the evolution of deuterium
after BBN, in order to correctly infer the primordial abundance
from observations in the $z \ll 10^{10}$ universe. 

Key to D is that there is no significant astrophysical
production site except for the big bang,
and that stars destroy D in their fully convective, pre-main sequence
phase \citep{bodenheimer}.
Together, these guarantee that any measurement of
D is a solid lower bound on the primordial 
abundance, and that in sufficiently primitive environments
D should be essentially primordial.
The lack of astrophysical D production 
was established in the classic paper by \citet[hereafter ELS]{els},
who considered all known sites in which
{\em any} nucleosynthesis occurs, and
demonstrated that none of these produce D in
significant quantities.

Deuterium is observed in diverse astrophysical settings.
In high-redshift ($z \sim 3$) QSO absorption systems,
neutral D is observed. The 5 best systems 
\citep{bt98a,bt98b,omeara,kirkman,pb}
give
$({\rm D/H})_{\rm QSOALS} = (2.78 \pm 0.29) \times 10^{-5}$.
The D abundance in the solar nebula is inferred from
solar wind observations of \he3 (which measure pre-solar D+\he3)
minus meteoritic determinations of \he3 alone.
These 
give \citep{gg} a value
$({\rm D/H})_{\rm pre-\odot} = (2.1 \pm 0.5) \times 10^{-5}$,
which probes proto-solar material 4.6 Gyr ago, or
at $z \sim 0.4$.
D/H is also observed \citep{ry,linsky}
in the local interstellar medium (ISM); 
recent FUSE observations \citep{moos} give
$({\rm D/H})_{\rm ISM} = (1.52 \pm 0.08) \times 10^{-5}$
in the Local Bubble today, at $z=0$.

The central ELS argument--that the big bang is the sole source of D--has 
stood the test of time, borne out by the drop in D/H
from its high-redshift to pre-solar to local ISM abundance.
But given the crucial role of D in cosmology,
it is important to carefully examine the assumptions
made, and to identify any possible loopholes.
Moreover, such an effort becomes crucial in light of
growing evidence that D/H has large variations over
short distances in the local ISM.
Recent FUSE observations \citep{hoopes}
add weight to earlier suggestions \citep[and refs.\ therein]{alfred}
that D/H can vary by as much as a factor of 2
over different lines of sight, pointing to
inhomogeneities on scales $\ga 100 \ \rm pc$.\footnote{
More observations are needed to 
firmly establish the nature and degree of D variations, and their
(anti-)correlations with metallicity \citep{gary}.
Nevertheless, it is already clear that FUSE is opening new windows
on the Galactic and local evolution of deuterium; this in turn
motivates a careful re-examination of the basic ELS assumptions 
which have guided our thinking on these issues.
}

Such a re-analysis of ELS was recently carried out by
\citet[hereafter ML]{ml}.
Specifically, they discussed D production by
suprathermal energetic particles in flare
stars.
Although ELS had considered flare production, 
ML note that ELS had neglected flare
production of neutrons, which can then go on
to produce D via radiative capture onto a proton
\beq
n + p \rightarrow d + \gamma
\eeq
with the emission of a photon with energy $E_\gamma = 2.223$ MeV.
ML do not make a detailed calculation of $n$ yields from
spallation reactions, but note that $n$ production can
be large compared to direct $d$ production, depending on the energetic
particle spectra.
ML then suggest that D created in such a way can escape from the flare site 
into the stellar wind.
Also, because of its low mass 
D might be preferentially ejected
relative to the heavier Li, Be, an B
which would otherwise contaminate the ISM. 
ML thus raise the possibility that flares could be a
significant source of D in the Galaxy.
From the point of view of Galactic chemical evolution,
this mechanism might then explain some or all of the ISM
variation in D.
Moreover, from the point of view of cosmology,
the chance of significant D production would
call into question the ELS argument
which underlies the recent spectacular agreement between high-redshift
D/H measurements and the predictions of BBN with the WMAP baryon density.

Thus, the ML scenario has important implications
and deserves careful further examination.
In this paper, we expand the analysis of ELS to
include the $n$ channel for D production suggested by ML.
Our approach is to identify general
nuclear physics constraints that follow from 
detailed calculations of $n$, $d$, and $\gamma$-ray production
by energetic particles.
The result is to close this loophole--i.e.,
we show that flare D production is not sufficient to offset
the monotonic decline of D with time.
However, flare D production should occur at {\em some} level,
and we find 
that $\gamma$-ray observations by the
recently launched INTEGRAL mission
may be able to probe this process.

\section{Light Element Production by Flares}

Flares are violent releases of magnetic energy during which
matter reaches temperatures of tens or hundreds of million Kelvin, and
particles get accelerated to very high energies. They are
associated with active stellar regions \citep{flares}. Nuclear
reactions then occur between flare-accelerated particles and
the ambient stellar atmosphere. In this section we will
discuss spallation reactions that result in production of
neutrons, deuterium and lithium. We will show that, as a
result of spallation, $1 \la n/d \la 10$ for most spectral indices, 
which means that ML channel
is indeed {\em more} important than channels considered by ELS, 
confirming and quantifying the suggestion of ML.
However, we will see that even including this channel for 
$d$ production, the concomitant Li production remains large
and thus severely constrains the possibility that flares
are an important source of D.

\subsection{Formalism}

Nuclide production in flares
depends on the initial spectrum of the flare,
its modification as it interacts with the stellar atmosphere,
and the compositions of the flare and the atmosphere.
In general, the nucleosynthesis yields in flares
have a complicated time dependence.
Fortunately, there are
two limiting approximations for  energetic
particles in flares: the thin-target and thick-target models
\citep{Ramaty75}. In the thin-target model, particle production 
is assumed to occur before the 
spectrum of accelerated particles is modified 
by ionization energy losses.
On the other hand, in the 
thick-target model, energy losses are taken to be large,
and particles slow due to ionization losses as
they move downward form the flare region,
prior to nuclear interactions.

But are both of these scenarios equally important for flare processes? 
A simple analysis of the mean free path of projectile particles sheds
light on this question. 
The mean free path of projectiles against nuclear interactions is
\begin{eqnarray}
\lambda = \frac{1}{n \sigma}
\end{eqnarray}
where $n$ is the target number density and $\sigma$ is the reaction cross section,
while mean free path of a flare particle (charge $Z$, mass number $A$)
that is losing energy due to its 
interaction with the surrounding medium is given by
\begin{eqnarray}
\lambda_{\epsilon} = \frac{A}{Z^2} \frac{R_p(\epsilon)}{n m_p}
\end{eqnarray}
Here, $\epsilon$ is the projectile energy per nucleon,
$n$ stands for the number density of the medium, 
$\sigma $ is the cross-section for a reaction of interest, 
$m_p$ is the proton mass, 
and $R_p$ is the ionization energy loss range of protons 
in units of $\rm g/cm^2$. 
In the thin-target limit $\lambda \ll 
\lambda_{\epsilon} $ which puts a lower limit on the projectile range
\begin{eqnarray}
R_p (\epsilon) \gg 167 \ \frac{Z^2}{A} \ 
{\left( \frac{\sigma}{10 \rm mb} \right) }^{-1}  \,\,\,\,\,\, 
\frac{ \rm g}{\rm cm^2}
\label{eq:R-lim}
\end{eqnarray}
For low energies, 
$R_p(\epsilon) \simeq 4 \times 10^{-4} \ {\rm g/cm^2} \ 
	(\epsilon/{\rm MeV})^2$,
so that eq.\ (\ref{eq:R-lim}) implies that 
the thin-target approximation holds for energies
\beq
\epsilon \ga 600 \ \sqrt{Z^2/A} \ \  \rm MeV/nucleon
\eeq
This energy is much larger than those of typical flare particles.
Thus, we conclude that the thick-target approximation is well-satisfied
and is the appropriate one for flare processes.  
This result, based on the physics of particle propagation,
is in agreement with the empirical result of \citet{thick}
who found that flare data are best described by thick-target rather than
thin-target models.

In the thick target model,
the production of secondary particles of species
$l$ from reaction $i+j \rightarrow l+ \cdots$ is given by \citep{Ramaty75}:
\beq
Q_l =  y_i y_j \int_{\epsilon_{\rm th}}^{\infty} \
\frac{\sigma_{ij}^l(\epsilon)}{\langle m \rangle}
\frac{dR_i}{d \epsilon} d \epsilon \int_{\epsilon}^{\infty} N_p (\epsilon ') d \epsilon ' 
\eeq
where $Q_l$ is the total number of secondary particles produced
per incident flare particle, and $\epsilon_{\rm th}$ is the
threshold energy for a particular reaction.
The ambient number density of target-species $j$ is $n_j$ and
$\sigma_{ij}^l$ is the cross section for the reaction of interest.
Projectile and target abundances are
given as numbers relative to hydrogen:  $y_i = n_i/n_p$; 
these are used to compute the mean target mass
$\langle m \rangle \simeq m_p \sum_i A_i y_i/\sum_j y_j = 1.4 m_p$. 

The projectile and target compositions 
are poorly known for flare stars, but one can imagine
several possibilities.
Since solar flares show enhanced metals and helium relative to
hydrogen \citep{mrk}, one might expect similar enrichment for
flare stars.
On the other hand, one expects M dwarfs in general to have
metallicities similar to those of G or K dwarfs 
\citep{wg,kfcm},
and thus on average to be subsolar in metallicity.
Thus we will examine several possible
compositions and find the impact of composition on
spallation production.
For a fiducial case, we will
assume solar abundances \citep{abundances}
for both projectiles an targets.
We then will examine the effect of more extreme variations:
a five-fold increase in metals and helium as in solar flares,
and a metal-free primordial case.

The spectrum $N_i(\epsilon) \ d\epsilon$ 
measures the total number of projectile
particles with energy in $(\epsilon,\epsilon+d\epsilon)$.
Following \citet{Ramaty75}, we adopt
the power-law form
\begin{eqnarray}
\label{eq:spec}
N_p(\epsilon ) &=& k_p \epsilon^{-s} \\
N_i(\epsilon ) &=& y_i N_p(\epsilon )
\end{eqnarray} 
where $k_p$ is the constant determined by normalizing $N_p$ to 1
proton of energy greater than $\epsilon_0=30$ MeV and $s$ is the
spectral index which is assumed to be the same for all accelerated
particles. The choice of this particular spectrum normalization is
purely conventional (see e.g. \citet{Ramaty75}), but in any case it
will not matter in our analysis since we are only interested in ratios
of productions of different particle species, and the normalization
will drop out.

In addition to the particle compositions, 
the spectral index $s$ is the other key parameter 
for the results we present.  We will present results
for a range of $s$, but some observations exist which 
constrain its value.
For solar flares, 
\citet{rmk} find that $\gamma$-ray line ratios
in an ensemble of flares point to a range of spectral indices
of about $s = 4 \pm 1$. 
For flare stars themselves, 
information on the spectra of the ions is not available,
but radio data from electron synchrotron emission
points to {\em electron} spectra with $s_e = 2$ to 3.
Whether the ions should have the same indices is, however,
unclear.
In the case of the Sun,
\citet{lmh} find that electron spectral indices (for $E_e < 100$ keV)
are correlated with proton indices, but
with $s_e = 2$ corresponding to $s \simeq 4$; they find
no significant correlation for $E_e > 200$ keV.
These results hint that the ion spectral indices in 
flare stars may also prefer $s \simeq 4$, but 
this parameter remains uncertain.

The particle ionization energy loss is measured by 
\begin{equation}
\frac{d \epsilon}{dR} = \frac{|d\epsilon/dt|}{\rho v}
\end{equation}
with units of [MeV/(${\rm g \ cm^{-2}}$)] where
$v$ is the velocity of
incident particle. 
We adopt the usual Bethe-Bloch energy formula
\citep{bethe} 
\begin{eqnarray}
\frac{{d \epsilon}}{dR} = \frac{4 \pi z Z^2 e^4}{A \langle m \rangle m_e v^2} 
\left[ \ln \left( \frac{2 \gamma ^2 m_e v^2}{I} \right) - \frac{v^2}{c^2} \right]
\end{eqnarray}
In this equation $Z$ is the charge of projectile, $z$ is number of
electrons per atom which is approximately equal to 1 in the solar
atmosphere, $A$ is the number of nucleons of projectile, $m_e$ is
electron mass, and the mean excitation potential is the value 
for hydrogen, $I=13.6$ eV.

Table 1 lists the reactions we have included, 
their lab-frame threshold energies,
and references for the 
cross-sections.
We note here that
cross sections for deuterium production via $\alpha + \alpha
\rightarrow D + \cdots$ are unavailable. To make an
estimate of deuterium production via $ \alpha \alpha$ reactions we
use exponential fit of \citet{Mercer} to the cross sections  for
$\alpha + \alpha \rightarrow d + \li6$, the channel
with the lowest threshold, and ignore other possible final states
($4d$, $dd\he4$). 
Also, the only existing cross sections for $\alpha +$C and $\alpha +$O production of 
deuterium are given for the single energy
$E_{\alpha} =58 {\rm MeV}$, or $\epsilon_{\alpha} =14.5$ MeV/nucleon 
\citep{bpk}. In this case we approximated the cross sections with:
\begin{equation}
\sigma(\epsilon) = 
  \left\{
  \begin{array}{cc}
    \sigma_0 \ e^{-2 \pi \eta} &  \epsilon > \epsilon_{\rm th} \\
    0 & \epsilon \le \epsilon_{\rm th} 
  \end{array}
  \right.
\end{equation}
where $\sigma_0$ is normalized to experimental values \citep{bpk},
and the Gamow factor $\eta = {z_i z_j e^2}/{\hbar c \beta}$
accounts for the Coulomb barrier.

\begin{table}
\begin{center}
\caption{Reactions included in our analysis.}
\begin{tabular}{ccc}
\tableline\tableline
Reaction & $\epsilon_{\rm th}$ & Reference 	\\
  & [MeV/nucleon] &  	\\

\tableline

p+p $\rightarrow$  n+x  & 410.20	& \citet{Ramaty75}	\\
p+ $\alpha \rightarrow$ n+x & 22.54 & \citet{Ramaty75} \\
p+CNO $\rightarrow$ n+x & 2.43 & \citet{Ramaty75}   	 \\
$\alpha + \alpha \rightarrow $ n+x & 9.44 & \citet{Ramaty75} \\
$\alpha $ +CNO $\rightarrow $ n+x & 0.49 & \citet{Ramaty75}	\\

&&\\

p+ $\alpha \rightarrow $ d+x & 23.07 & \citet{meyer} 		\\
p+CNO $\rightarrow $ d+x & 17.50 &  \citet{meyer}	\\
$\alpha + \alpha \rightarrow $ d+$^6$Li & 15.38 & \citet{Mercer}	\\
$\alpha $ + C $\rightarrow $ d+x & 4.53 & \citet{bpk}	\\
$\alpha $ + O $\rightarrow $ d+x & 5.10 & \citet{bpk}	\\
 
&&\\

p+C $\rightarrow ^6$Li+x & 24.45 & \citet{rv}	\\
p+N $\rightarrow ^6$Li+x & 17.52 & \citet{rv}	\\ 
p+O $\rightarrow ^6$Li+x & 23.57 & \citet{rv}	\\ 
$\alpha + \alpha \rightarrow ^6$Li +x & 11.19 & \citet{Mercer}	\\
$\alpha $+C $\rightarrow  ^6$Li +x & 7.91 & \citet{rv}	\\ 
$\alpha $+N $\rightarrow  ^6$Li +x & 2.83 & \citet{rv}	\\ 
$\alpha $+O $\rightarrow  ^6$Li +x & 6.02 & \citet{rv}	\\ 

\tableline
\end{tabular}
\end{center}
\end{table}

\subsection{Results}

\begin{figure*}[t]
\epsscale{0.55}
\plotone{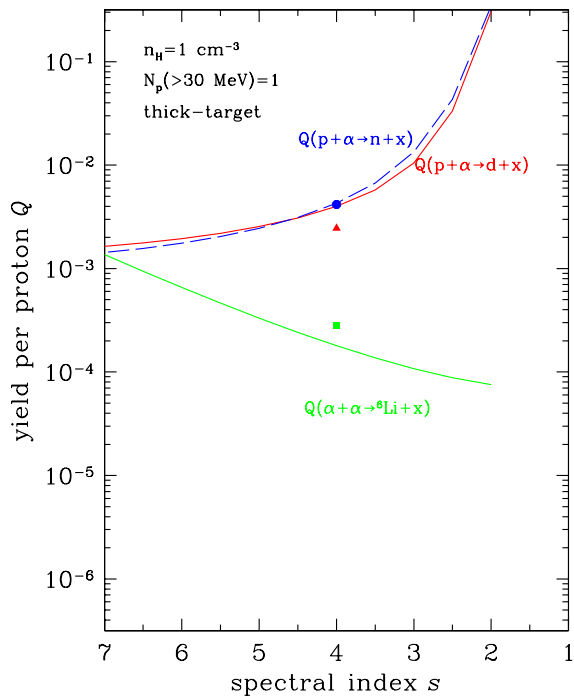}
\caption{
Total yields of  neutrons, deuterium and $^6$Li
produced via spallation in the thick-target model,
in particles produced per flare proton above 30 MeV. 
For clarity, we show only the reaction that dominates production
of each element over most of the range of $s$.
The points 
at $s=4$ represent
our analytic estimates for this choice of spectral index;
$n$=circle, $d$=triangle, $\li6$=square.
Abundances used is $y_{\rm He}
=0.1$. Threshold energies and flat cross-sections were approximated by
$\epsilon_{\rm th} \approx 25$ {\rm MeV} and $\sigma_{p \alpha}^d \approx 40$ 
{\rm mb}
for deuterium production, $\epsilon_{\rm th} \approx 22$ {\rm MeV} and $\sigma_{p \alpha}^n \approx  60$ for neutron production,  $\epsilon_{\rm th} 
\approx 11$ {\rm MeV} and
$\sigma_{\alpha\alpha}^{^6{\rm Li}} \approx 20$ {\rm mb} for \li6.
}
\label{fig:check}
\end{figure*}

\begin{figure*}[t]
\epsscale{0.55}
\plotone{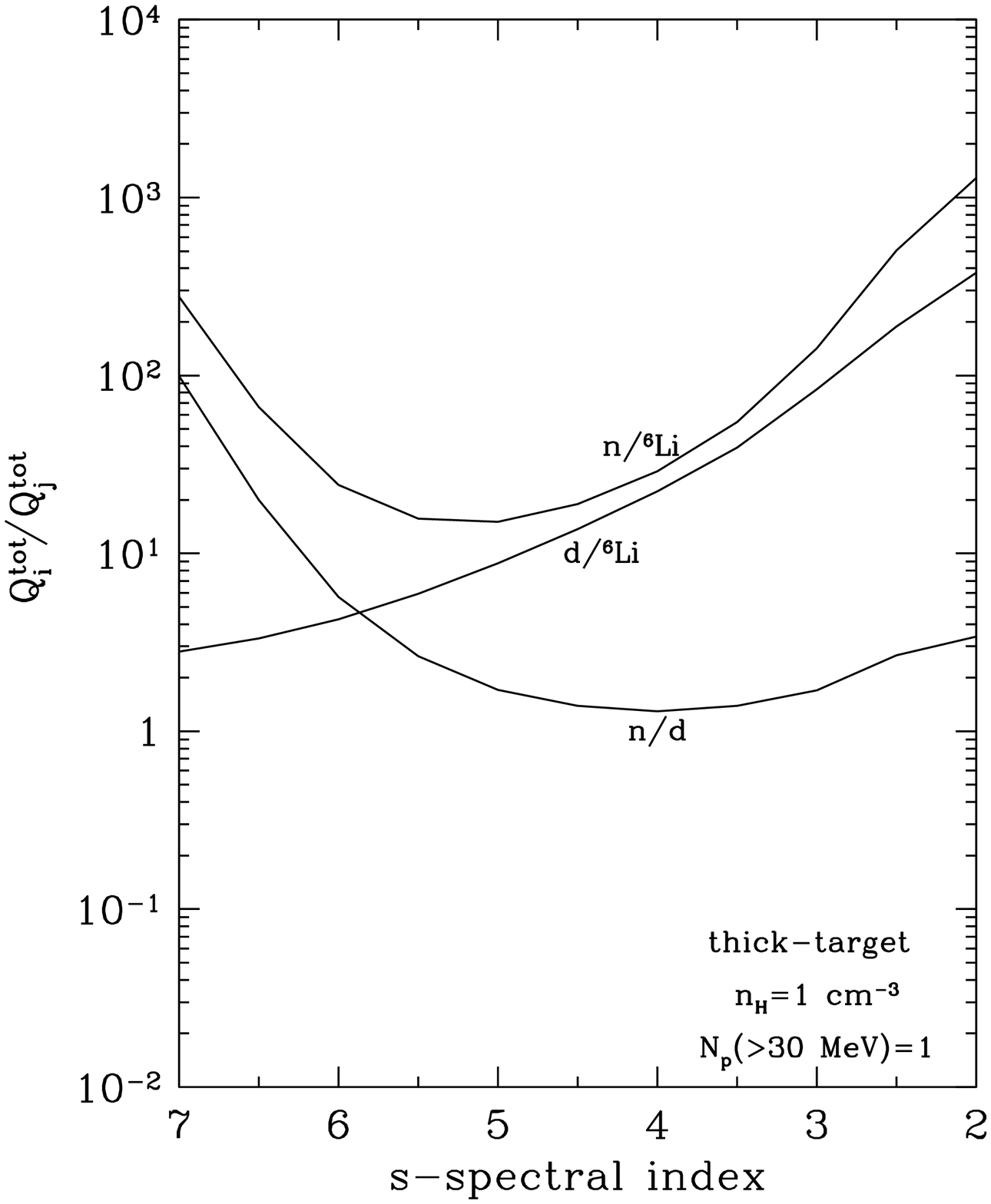}
\caption{
The ratios of total numbers of secondary particles 
produced via spallation in the thick-target case. All ratios are shown
as functions of the spectral index s. }
\label{fig:results}
\end{figure*}

Before we present our final results let us first make analytic
estimates for couple of reactions for this model. In order to perform analytic estimates some approximations
will be needed. 
To analytically solve thick-target case we
approximate energy losses via:
\begin{equation}
 \frac{dR}{d \epsilon} 
  =  0.14
   \frac{A}{2 \pi Z^2 z} 
   \frac{m_e \langle m \rangle}{m_p e^4} \ \epsilon
\end{equation}
Let us also assume that a
cross-section for a particular reaction is flat and that we
can neglect relativistic corrections which is a valid assumption for
low projectile energies.
Therefore, equation (6) now has the following form:
\beq
Q_l = 5.7 \times 10^{-4} \frac{A  y_i y_j}{(s-3)Z^2 } 
   \left( \frac{\sigma}{10 \, {\rm mb}} \right) 
  { \left( \frac{ \epsilon_{\rm th}}{ 10 \, {\rm MeV}} \right) }^2 
  { \left( \frac{\epsilon_{\rm th}}{\epsilon_0} \right) }^{1-s}
\eeq

In Figure \ref{fig:check}
we present our analytic estimates along with numerical
results for deuterium production by $p \alpha$ ,\li6 production
by $\alpha \alpha $ and neutron production by $p \alpha$ reactions. The analytic result is calculated for the case of
$s=4$ which is the best choice given the assumptions that were made,
while numerical results are functions of spectral index $s$. Note that
our results also include reactions where target and projectile species 
are interchanged (initial-state particle species can also serve
as targets, since we assume that flares have the same
composition as the ambient medium). 
Although we made some rough estimates we see that they
are in good agreement with our numerical predictions, which gives
us confidence in our code.

Our numerical results for production of neutrons, deuterium and
\li6 are shown in Figure \ref{fig:results} .
Results are represented in the form of  ratios of total numbers of particles produced in
spallation as a function of the spectral index $s$.  The purpose of this was to see how neutron production
compares to production of deuterium, since the key idea of ML
paper was that a significant amount of neutrons can be produced via
spallation which can then undergo radiative capture to make a
non-negligible amount of deuterium. 
We see that indeed, $n/d \ga 1$, so that neutron production
does in fact dominate direct $d$ production for all spectral
indices $s$.  More specifically, $n/d \sim 1$ around
$s = 4$, the best-fit value for solar flares, and
can be as high as $n/p \sim 200$ for an extreme value of
$s = 7$.
These calculations thus confirm the suggestion of ML
that in fact $n$ production is significant compared to
deuteron production, and thus radiative capture of the neutrons
offers an important channel for deuterium synthesis which
had been neglected by ELS.

With the neutron production in hand, we can now update the
ELS argument to address the ML loophole.
The key point here is that spallation production of 
$d$ and $n$ is also inevitably accompanied by 
production of other light elements, notably lithium.
In particular, since \li6 is 
uniquely produced by spallative processes \citep{fo,elisa},
it offers the strongest constraint, as follows.
Assume that all neutrons made in flares undergo radiative capture onto
protons,
rather than suffering decay or non-radiative capture on \he3. 	
We also assume that all \li6 in the ISM comes
from flares, which maximizes the flare contribution.
Then by using our
$(n+d)/\li6$ ratio for the entire range of spectral indices,
$20 \la (n+d)/\li6 \la 1000$, 
combined with the solar $\li6/{\rm H} = 1.5 \times 10^{-10}$, 
we infer a flare-produced deuterium abundance in the range
$3 \times 10^{-9} \la {\rm D/H} \la 1.5 \times 10^{-7}$,
much smaller than observed deuterium abundance. 
This limit would be further strengthened if we also note 
that a large fraction ($\ga 50\%$) of neutrons
escape the Sun, or 
are captured onto \he3 rather than on protons \citep{Reuven}.
If $f_{np} \la 0.5$ is the fraction of $n$ which do capture
on protons, 
thus the correct ratio for D/\li6 is 
$(f_{np} n+d)/\li6 = (f_{np}n/d+1)d/\li6$, which 
lowers the total deuterium production by
an additional $\sim 25\%$ for $n/d \sim 1$ and $f_{np} = 1/2$.

We can also go the other way. Assume that just 10\% of observed ISM
deuterium abundance comes from flares. Then from our $(n+d)/\li6$ ratio for
all spectral indices we get that \li6/H abundance is in the range 
$1.5 \times 10^{-9} \la \li6/{\rm H} \la 7.5 \times 10^{-8}$,
which is between 1 and 2.5 orders of magnitude larger than 
the observed solar abundance of \li6 (where overproduction of
about $300$ corresponds to the spectral index most favored for solar flares). 
Thus we conclude that then we can rule out the ML loophole,
if flare spallation products escape with equal 
probability.  This result updates and reaffirms the same
conclusion by ELS.

\begin{figure*}[t]
\epsscale{0.9}
\plottwo{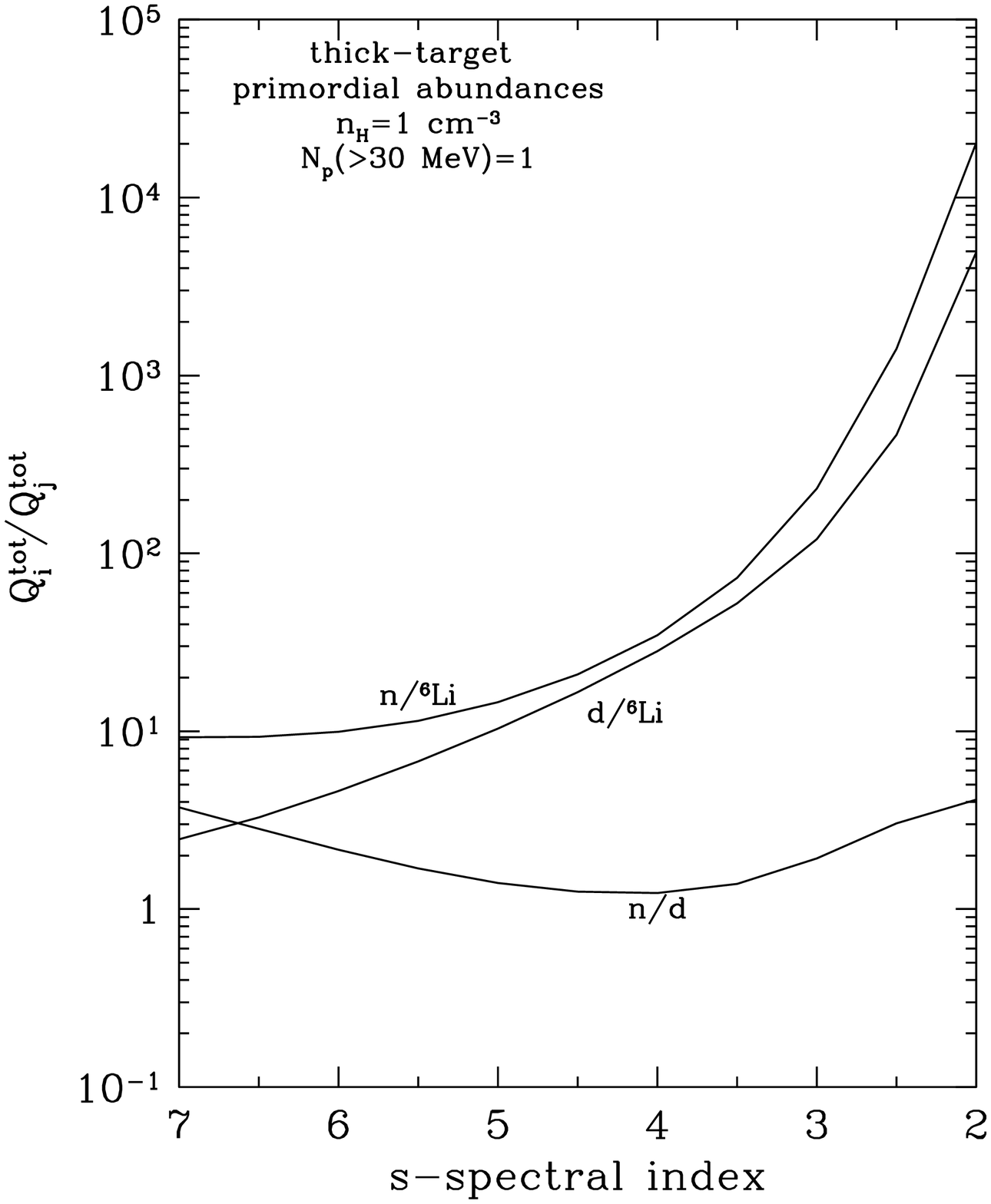}{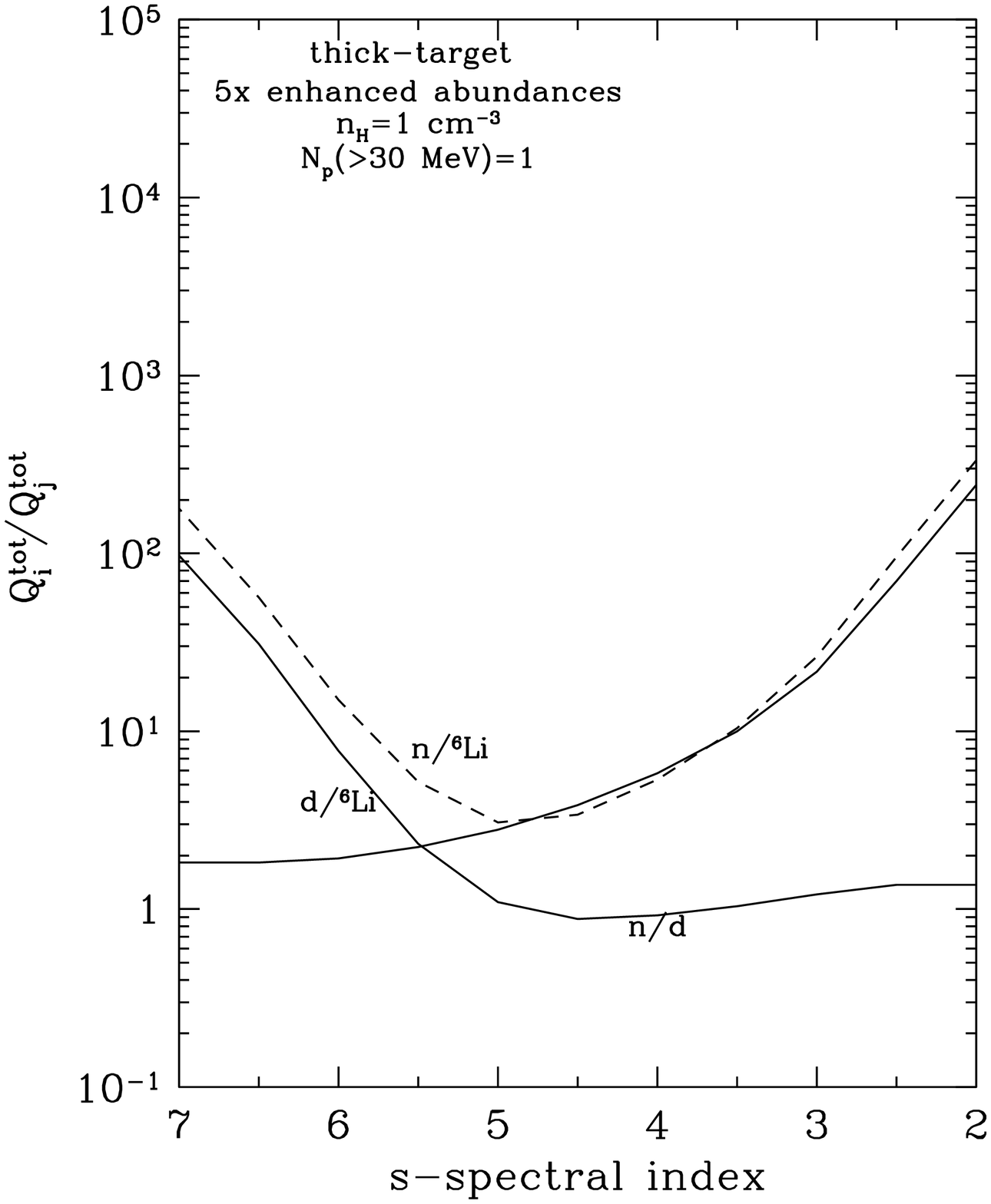}
\caption{
As in Figure \ref{fig:results}, but
with projectile and target abundances as follows:
(a) primordial, i.e., He/H = 0.1, no metals; and
(b) metal-enriched, as is seen in solar flares. 
Here we take
$y_{\alpha}=0.5$, $y_{\rm O}=4.24 \times 10^{-3}$, $y_{\rm N}=5.60
\times 10^{-4}$ and $y_{\rm C}=1.79 \times 10^{-3}$.
\protect{\label{fig:altabs}}
}
\end{figure*}

However,
many flare stars will have fewer metals, and thus a different composition.
For example, disk G and K dwarfs are known \citep{wg,kfcm}
to have a mean metallicity that is subsolar by about a factor of 2.
What effect will this have on the production ratios and ultimately 
the revised ELS constraints? 
To get a sense for this, we follow ELS and repeat our calculation for
the extreme limit of
a {\em primordial} composition--i.e., where the flare
and the ambient medium contain only \he4 (with
mass fraction 24.8\%) and the balance is H.  
The difference here is that we exclude
reactions involving CNO, but since Li can be made by $\alpha \alpha$ fusion,
the \li6 constraint remains.  In this case the 
$(n+d)/\li6$ ratio is in the range of 
$10 \la (n+d)/\li6 \la 30000$, 
as shown in Figure \ref{fig:altabs}a, 
thus placing the upper limit (for very low spectral indices) on the
expected deuterium abundance due to flares about an order of magnitude
higher. Therefore we see that even in primordial environment the ELS
constraint holds, especially since thick-target production ratios that
we are discussing change only slightly around most favorable spectral
index of $s \approx 4$.

Now, although our fiducial calculation 
is based on the solar abundances, 
the Sun itself shows a different and more enriched composition
in flare projectiles and possibly targets.
In solar flares, \he4/H can be as much as 5 times higher than
in mean solar matter--in which case protons and $\alpha$ particles
can be within a factor of two of each other by number;
abundances of other heavy elements can also be similarly enhanced in flares 
\citep{mrk}. 
Therefore, one might wonder how will our results change in that case? 
Figure \ref{fig:altabs}b plots our results obtained as before,
but this time for abundances that are 5 times greater than
solar. However, even in this case our main conclusions stay the same:
({\em i}) the neutron channel for D production can dominate over
spallation production of D, but not enough to be a significant source
of non-primordial D, and  ({\em ii}) the ELS constraint on
\li6/D holds, and indeed
is the strongest in
this case, since overproduction of \li6 goes between 1 and 3 orders of
magnitude. Therefore we see that by using a solar composition to
describe flare-processes, our constraints were in fact generous
compared to this case.

Of course, ML note that it is possible that D escapes more
readily than does Li, due to its small mass.  
Clearly, the question of transport is complex, involving
competing effects such as convection and mass loss.   
These difficult issues in magnetohydrodynamics
are beyond the
scope of this paper, but we can at least quantify
the bias needed to avoid our updated ELS constraint.
We saw that if interstellar D is due to flares, 
then \li6 is overproduced by a factor that is about between 10 and 250.
Thus, \li6 mass loss must be suppressed relative to D by
at least this factor in order for the ML loophole to remain open.
While it is difficult to rule this possibility out (or in!), clearly
this offers a strong quantitative constraint on the
particle loss mechanism.

Thus, we strongly constrain the ML loophole, which
now requires a strong bias against Li escape.
But if such a bias could be created, is the ML loophole viable?
To further explore that scenario, we
turn to the 2.22 MeV $\gamma$-ray line as an additional constraint on
neutron captures in the Galaxy.

\section{Gamma-Ray Line Constraints on Flare Production of Deuterium}

In the previous section we have seen that analysis of particle
production through spallation reactions does not completely answer the
question of the amount of deuterium produced in stellar flares. For
that reason we will now approach this problem from a different angle.
The reaction of radiative capture
$ n + p \rightarrow d + \gamma $
which was proposed to be potentially significant source of deuterium
in ML paper, has as a result, besides deuterium, a $2.22$ MeV
$\gamma$-ray line. Our idea is to predict the Galactic $\gamma$-ray
intensity of that line under the assumption that 
flares produce significant
amounts of deuterium. That way, without going into details about
mechanisms that can transport produced deuterium into the ISM, we can
place an upper limit on non-primordial deuterium production via
radiative capture.

If we denote number of 2.22 MeV $\gamma$-ray lines produced by a
single flare with $N_{\gamma}$, and use
$N_d^{\rm tot}$ to denote number of deuterium produced by the same flare
in both radiative capture and spallation processes then
we can write
\begin{eqnarray}
\frac{N_d^{\rm tot}}{N_{\gamma}} = \frac{1}{f_{2.2}} \frac{N_d^{\rm tot}}{N_n} 
\end{eqnarray}
where $N_n$ is the number of neutrons produced by the same star via
spallation. The factor $f_{2.2}$ is the efficiency
for neutron-to-2.22 MeV photon
conversion, averaged over neutron energy spectrum
\citep{Reuven}; this includes stopping of the $\gamma$ rays,
as well as neutron escape and neutron ``poisoning'' by capture onto \he3. 
Since deuterium is made via radiative capture as well
as in spallation, we have 
$N_d^{\rm tot} = N_d^{\rm spall} + N_d^{\rm rc} $.
Let us further assume
that all neutrons made in spallation reactions, $N_n$, go into making of
deuterium by radiative capture, that is, assume that $ N_d^{\rm rc}= N_n$.
Thus the 2.22 MeV $\gamma$-ray intensity estimated with this assumption
corresponds to the upper limit of Galactic flare production of
deuterium. We can now rewrite equation (14) and obtain $
\ndot_d^{\rm tot} $, the rate of total deuterium production by a
single star, as a function of deuterium-to-neutron spallation production
ratio
\beq
{\ndot}_d^{\rm tot} = \frac{{\ndot}_{\gamma}}{f_{2.2}}
\left[ 1+ {\left( \frac{d}{n} \right)}_{\rm spall} \right] 
\eeq

What we actually want is the Galactic deuterium production rate by
mass, $ {\dot{M}}_d^{gal} $, which is a function of total deuterium
production rate ${\ndot}_d^{\rm tot}$  and total number of flare stars in
the galaxy $\ndme$:
\begin{eqnarray}
{\dot{M}}_d^{\rm gal} &=& m_d {\ndot}_d^{\rm tot} {\cal N}_{*}^{\rm gal} \\
 &=& \frac{{\ndot}_{\gamma}}{f_{2.2}} 
  \left[ 1+ {\left( \frac{d}{n} \right)}_{\rm spall} \right] 
  m_d {\cal N}_{*}^{\rm gal}
  \label{eq:Mdot}
\end{eqnarray}
where ${\cal N}_{*}$ is the number
of flare stars in the Galaxy, and
$m_d$ is the deuteron mass.

We wish to find the observable Galactic 2.22 MeV $\gamma$-ray intensity
$I_{\gamma} $, 
in terms of the 2.22 MeV emissivity $ {q}_{\gamma}^{\rm gal}$,
the  total
Galactic production rate of 2.22 MeV $\gamma$ rays per volume
per steradian
\beq
\label{eq:emissivity}
{q}_{\gamma}^{\rm gal} = \frac{{\ndot}_{\gamma} n_{*}}{4 \pi}
\eeq
where $n_{*}(\vec{r})$ is the number density of flare stars.
The intensity is simply a line-of-sight integral:
\beq
I_{\gamma} 
  = \int_{\rm los} {q}_{\gamma}^{\rm gal} \ dl 
  = \frac{{\ndot}_{\gamma} N_{\rm los}}{4\pi}
\eeq
where $N_{\rm los} = \int_{\rm los} n(\vec{r}) \ dl$ is the column 
density of flare stars along the line of sight $l$.

Therefore, we can link
Galactic deuterium production rate by mass as a function of Galactic
2.22 MeV $\gamma$-ray intensity
by combining eqs.\ \ref{eq:Mdot} and \ref{eq:emissivity}
\begin{eqnarray}
\label{eq:D2gamma}
{\dot{M}}_d^{\rm gal} 
  = \frac{4 \pi}{f_{2.2}} 
\left[ 1 + {\left( \frac{d}{n} \right)}_{\rm spall} \right]
\frac{m_d {\cal N}_{*}}{N_{\rm los}} I_{\gamma}
\end{eqnarray}
or
\beq
\label{eq:gamma2D}
I_{\gamma} = 
\left[ 1 + {\left( d/n \right)}_{\rm spall} \right]^{-1} 
\frac{ f_{2.2} N_{\rm los}}{4 \pi m_d \ndme}  \dot{M}_d^{\rm gal}
\eeq
Equation (\ref{eq:D2gamma})
allows us to estimate the most modest Galactic 2.22 MeV
$\gamma$-ray intensity that we expect if the flare stars are
producing significant amounts of deuterium. 
In order to predict the lower intensity limit, let us
take $d/n = 1$ (although from Fig.\ 2
we see that
this ratio is always smaller then 1). 
We also adopt the $f_{2.2}=0.1$ neutron-to-$\gamma$ conversion
efficiency calculated by 
\citet{Reuven};
this value is dominated by escape (followed by decay), and
then the neutron poisoning
due to the large $n$-capture cross section for
\he3.

The Galactic distribution $n(\vec{r})$ of flare stars (i.e., M dwarfs)
controls both the column density and flare star number
which appear in eqs.\ (\ref{eq:gamma2D}) and (\ref{eq:D2gamma}).
We have verified that disk M dwarfs dominate the calculation,
while spheroid (``stellar halo'') M dwarfs \citep{gfb}
do not contribute
significant numbers or column of stars.
For the disk M dwarfs, we adopt the \citep{zheng} distribution
\beqar
\label{eq:Mdist}
n(r,z) & = & n_0 \ \exp\left(-\frac{r-R_0}{H}\right) \times \\
\nonumber
 & &  \left[ (1-\beta) \exp(-|z|/h_1) + \beta \exp(-|z|/h_2) \right] 
\eeqar
\citet{zheng} use {\em Hubble Space Telescope} star counts
to determine the initial mass function in the M dwarf range,
and find the parameters in eq.\ (\ref{eq:Mdist}) to be:
local M dwarf density
$n_0 = \rho_0/\langle m_{\rm M \, dwarf} \rangle 
  = 5.2\times 10^{-2} \ {\rm pc^{-3}}$,
radial scale length $H = 2.75\ {\rm kpc}$,
solar Galactocentric distance $R_0 = 8 \ {\rm kpc}$,
and vertical scale parameters
$h_1 = 156 \ {\rm pc}$,
$h_2 = 439 \ {\rm pc}$, 
and $\beta = 0.381$.
With these, we find the number of Galactic 
M dwarfs to be $\ndme = 2.4 \times 10^{10}$,
and a column density towards the Galactic center of
$N_{\rm los} = 5.1 \times 10^3 \ {\rm pc^{-2}}$.

The deuterium production rate that
we would consider to be significant is one that is comparable
to the Galactic destruction rate of deuterium due to
its cycling through stars (``astration''), 
which burn D during the pre-main sequence phase. 
The present rate at which deuterium is lost is 
\beq
{\dot{M}}_{d,{\rm astrate}} = -X_d {\cal E}
                = -2.5 \times 10^{-5} \, M_{\odot}/ {\rm yr}
\eeq
where 
$X_d = 2.1 \times 10^{-5}$ is the present ISM mass fraction of deuterium,
and ${\cal E}$ is the current rate at which mass is
ejected from dying stars.
This is roughly given by ${\cal E} = R \psi$,
where $R \simeq 0.3$ is a conservative estimate
of the ``return fraction'' of mass from stars \citep[e.g.]{pagel},
and we adopt a current Galactic star formation rate
$\psi \approx 4 \, M_{\odot}/{\rm yr}$
following, e.g., \citet{focv}.

We then can place a  lower limit on the 2.22 MeV Galactic $\gamma$-ray 
intensity that corresponds to flare production of deuterium equal to the 
amount of deuterium that is destroyed per year in the Galaxy:
\begin{eqnarray}
\label{eq:MLgamma}
I_{\gamma} 
 = 4 \times 10^{-2}  \ {\rm cm}^{-2} \ {\rm sec}^{-1} \ {\rm sr^{-1}}
\end{eqnarray}

We can compare this prediction with 2.22 MeV all-sky map done by
COMPTEL
\citep{Comptel}.  
This map contains a single possible 
point source but no diffuse emission is found, down to a level
estimated (M. McConnell, private communication)
to be 
\beq
I_\gamma^{\rm obs} 
  <  5 \times 10^{-4} \ {\rm cm^{-2} \, sec^{-1} \, sr^{-1}}
\eeq
This limit is about 1\% of what we would expect 
if a significant amount of deuterium originates in flares
(eq.\ \ref{eq:MLgamma}). 
Or to go other way around, we can use eq.\ (\ref{eq:gamma2D})
to translate the COMPTEL upper limit
into an upper limit for deuterium
production by flares:
\begin{eqnarray}
\label{eq:maxMdot}
{\dot{M}}_d^{\rm gal} \la 3.0 \times 10^{-7} \ M_{\odot}/{\rm yr}
 \simeq 1 \times 10^{-2} | {\dot{M}}_{d,{\rm astrate}} |
\end{eqnarray}
Thus we see that the maximum possible D production rate
is at most $1\%$ of the destruction rate, so that
the net effect is that D is indeed destroyed and thus
decreases with time.
If we were only to look at the $n$ channel of deuterium production, as it was proposed by ML, we would get even lower production rate 
(by about a factor of two). 
Therefore on the basis of COMPTEL observations and our numerical analysis of production rates we can conclude that stellar flares are 
{\em not} significant sources of non-primordial deuterium
on the Galactic scale, and thus do not spoil the
monotonic decline of D with time.

Although we have estimated the diffuse 2.22 MeV radiation
from flare stars, our calculation in fact is generally
applicable to any Galactic sources of deuterium via neutron radiative capture.
The $\gamma$-ray constraints are thus stronger than 
the abundance-based constraints of the previous section.
Furthermore, further limits (or observations) at 2.22 MeV
will constrain {\em any} source of D production from neutrons.
We encourage INTEGRAL observations to be made to tighten the constraints
we present here.

While $\gamma$-ray observations rule out flare D production at 
a level which would overwhelm destruction,
the fact remains that neutron production is a significant
and inevitable result of flare activity, as emphasized by ML.
Thus, a $\gamma$-ray signature of this process {\em must} 
exist at some level.
We can make a crude prediction of diffuse 2.22 MeV
$\gamma$-ray intensity that INTEGRAL might observe. The rate at which
2.22 MeV $\gamma$-ray are produced in a single star is given by
\beq
\label{eq:star_lum}
{\ndot}_{\gamma} = f_{2.2} \frac{Q_n}{\langle \epsilon \rangle} L_{\rm flare}
\eeq
where $L_{\rm flare}$ is the time-averaged luminosity in the flare state,
and 
\beq
\langle \epsilon \rangle 
  = \sum_i \int \epsilon \ N_i(\epsilon) \ d\epsilon
  = \int \epsilon k_p {\epsilon}^{-s} d\epsilon \sum_{i} y_i A_i
\eeq
measures the flare energy going into accelerated particles.
Using the spectral index $s=4$ favored by solar flares,
and integrating from the lowest threshold energy among reactions involved in neutron production (note that because of assumed power law spectrum of flare particles it is not possible to extrapolate to zero energy) we find that the number of neutrons produced per flare unit energy is
${Q_n}/{\langle \epsilon \rangle}=0.015 \ \rm atoms \ erg^{-1}$. 

For the total number of neutrons produced per proton (above 30 MeV),
we used our thick-target $s=4$ numerical result $Q_n=5.53 \times 10^{-3}$.
Then by taking a bolometric flare luminosity of dM5e star to be 
$L_{\rm flare}=1.9 \times 10^{29}$ erg/sec \citep{mdwarfs} and 
neutron-to-photon efficiency factor to be $f_{2.2}=0.1$ it follows 
from equation (19) and (20) that
\beq
\label{eq:flare2.2}
I_{\gamma} \sim 1.2 \times 10^{-8} \ \rm cm^{-2} \ s^{-1} \ sr^{-1}
\eeq
In this calculation we took $V_* = 10^{12} \ \rm pc^3$ and 
$n_{*}=0.86 \ \rm pc^{-3}$ \citep{weistrop}, and we
took the a line of sight length $l=20$ kpc towards the Galactic center. 
Finally, we note that using the same estimate of flare activity,
(i.e., scaling to the flare star luminosity),
we can also arrive at an estimate of the 
flare D production rate, namely 
${\dot{M}}_d^{\rm flare} \sim 2.5 \times 10^{-10} \ M_{\odot}/{\rm yr}$,
which is completely negligible compared to the 
D astration rate.

The emission in eq.\ (\ref{eq:flare2.2}) is low, 
and in particular is too dim to be observed by INTEGRAL.
On the other hand, our estimate is crude, and 
while detection of flare stars would be a surprise,
it would also provide unique new information
about the particle content and energetics in
stellar flares.
Also, as we have noted, simply
tightening the limits on diffuse 2.22 MeV radiation
will strengthen the case that D production is negligible, and
can go further to constrain (or probe) flares as a contributor
to the D variation in the ISM \citep{hoopes}.
For these reasons, we encourage INTEGRAL observations
at 2.22 MeV, particularly towards the Galactic center 
where the emission should be the strongest.\footnote{We note 
that COMPTEL \protect\citep{Comptel,Comptel2} found a single
point source at 2.22 MeV, with flux 
$\phi_\gamma = (3.3  \pm 0.9) \times 10^{-5} \ {\rm cm^{-2} \ s^{-1}}$.
A flare star with luminosity $\ndot_\gamma$ (eq.\ \protect\ref{eq:star_lum})
at distance $r$
has a time-averaged flux 
$\langle \phi_\gamma \rangle \simeq 3 \times 10^{-12} (r/1 \, {\rm pc})^{-2} 
 \ {\rm cm^{-2} \ s^{-1}}$.
Thus, the point source is too bright to be even a nearby 
flare star unless COMPTEL happened to see a brief, intense burst,
from a nearby star with flare luminosity much larger than the average dMe.
INTEGRAL observations of this source can test the flare star
hypothesis by looking for 2.22 MeV time variability in coincidence with, e.g.,
H$\alpha$ and other line emission, as well as X-ray and UV signatures.}

\section{Constraints on Localized Deuterium Production}

It is important to note, however, that this conclusion applies
to the Galaxy as a whole.  While the global D production rate
due to flares is small, it is a separate question whether
D production could be sufficient to create D variations on
smaller scales.

To constrain
local enrichment, we rely on the
Galactic average deuterium production rate $\dot{M}_d$,
which we have constrained 
in the previous subsection on the basis of 
gamma-ray observations, and estimated 
using flare energetics.
With the global production rate in hand,
we can find the average D production rate
for one star.
The limit from eq.\ (\ref{eq:maxMdot})
gives 
$\dot{m}_d = \dot{M}_d/\ndme \sim 10^{-17} \ \msol/{\rm yr}$. 
Using this, let us find the needed D production
to pollute a parcel of gas at a level of 
$\delta {\rm D/D} \sim \delta M_D/M_D \sim 1$,
i.e., to product a factor of 2 variation in D/H over the current ISM level.
To pollute a gaseous region of total mass $M_g$
given the entire age of the universe
requires that $M_{\rm M \, dwarf} \sim 70 M_g$,
i.e., a gas fraction of no more than $1/70$, 
that is, the M dwarfs must outweigh the 
ambient gas by a large factor.  On the other hand,
a clump of M dwarfs in the hot and diffuse ISM 
could achieve this level.
If we instead adopt the deuterium production rate 
$\dot{M}_d \sim 10^{-10} \ \msol/{\rm yr}$, based
on energetics, we find a needed gas fraction $< 10^{-5}$.
This becomes harder to achieve, but is difficult to completely exclude.

\section{Discussion and Conclusions}

For more than a quarter century, 
the ELS argument that all deuterium is primordial
has played a central role in cosmology.
The importance of this argument demands that
its assumptions be carefully checked.
In this context, Mullan \& Linsky's discovery
that D production by neutron capture in flares
can evade the ELS constraints is thus very important and
demands further investigation.

In this paper, we have examined  and constrained the ML scenario
in two ways.
First, we have made a detailed calculations of spallation yields
in flares.  These calculations show that
flares have $1 \la n/d \la 200$, confirming the ML suggestion
that neutron production is at least as significant as direct 
deuteron production, and considerably more for some spectral indices
(albeit ones atypical of solar flares).
We have thus updated the ELS constraints on flares, now taking into
account neutron production (all which we assume leads to
deuterium formation by radiative capture).  Specifically,
we consider \li6 production which accompanies $n$ and $d$
synthesis.  The ELS argument
is quantitatively changed,  because the $\li6/(n+d)$ ratio is lower than
the former $\li6/d$ ratio, but the qualitative conclusion remains,
that the solar \li6 abundance forbids a significant $d$ component from
flares.  The only way that flare nucleosynthesis could avoid this
bound is for $d$ to escape the star preferentially
with respect to \li6, the needed bias being a factor
around 1000 in escape efficiency.

Second, we considered the implications of the 2.22 MeV $\gamma$-ray
line produced in the radiative capture.  If D is currently produced
in the Galaxy via this mechanism (either in flare stars or elsewhere)
then there inevitably is an observable 2.22 MeV line signature.
In the case of flare stars, the emission would be unresolved and 
would lead to a diffuse intensity tracing the Galactic plane.  The
observed limits on 2.22 MeV emission from COMPTEL immediately
translate into a constraint on D production by radiative capture,
and rule out this mechanism as a significant source of D
at the global Galactic scale.

It is more difficult to constrain flare production of D
as the source of 
the possible scatter in local D abundances as
reported recently by FUSE \citep{hoopes}.
However, arguments based on energetics and on gamma-ray fluxes
do demand that the mean D production rate per star is
small.  This means either that (1) to produce significant fluctuations
requires a high local concentration of M dwarfs relative to the
gas they pollute, or (2) that flare D production is dominated
by a very small fraction of M dwarfs, which dominate the global mean
production and can thus create local anomalies.  
In either case, the nearest local sources may be 
detectable through their gamma-ray lines.

Another constraint on the ML scenario comes from observations of
deuterated molecules towards the Galactic center.
\citet{lubowich} detect DCN in a molecular cloud
10 pc away from the Galactic center, and
on the basis of a molecular chemistry model
estimate that  ${\rm D/H} = (1.7 \pm 0.3) \times 10^{-6}$
in this cloud.
Unfortunately, the fractionation corrections here are large and thus
the results are somewhat model-dependent.
With this caveat, 
the important point here is that D/H is found to be {\em lower}
towards the Galactic center, suggesting that D indeed has
a positive Galactocentric gradient, which argues against 
{\em any} significant stellar source of D, including flares.

Finally, we note that while flare star radiative capture synthesis of D
is insufficient to alter the conclusions of ELS, 
it is certain that the process does occur at some level--the
flares exist, and must produce neutrons.
Thus, the 2.22 MeV signature of this process must
exist at some level.  We have made a simple estimate of the
surface brightness towards the Galactic center. We find this to be
below the sensitivity of INTEGRAL, but given
the crudeness of our estimate, it is worth investigation,
as limits (or detection!) of this line has immediate implications
for stellar flares, 
neutron capture processes,
 and deuterium evolution in our Galaxy.

\begin{acknowledgments}

We thank Don York for discussions which stimulated this work.
We are particularly grateful
to Mark McConnell and Ed Chupp 
for very helpful discussions of the COMPTEL 2.22 MeV
results, and to Jeffrey Linsky and Dermott Mullan for constructive
comments on an earlier version of this paper.
This material is based upon work supported by the National Science
Foundation under Grant No. AST-0092939.

\end{acknowledgments}

\end{document}